# Reversible electron-hole separation in a hot carrier solar cell

S Limpert[1], S Bremner[1], and H Linke[2]

[1] School of Photovoltaic and Renewable Energy Engineering, University of New South Wales, 2052 Sydney, Australia

[2] NanoLund and Solid State Physics, Lund University, Box 118, 221 00 Lund, Sweden

E-mail: steven.limpert@student.unsw.edu.au, stephen.bremner@unsw.edu.au, heiner.linke@ftf.lth.se

**Abstract.** Hot-carrier solar cells are envisioned to utilize energy filtering to extract power from photogenerated electron-hole pairs before they thermalize with the lattice, and thus potentially offer higher power conversion efficiency compared to conventional, single absorber solar cells. The efficiency of hot-carrier solar cells can be expected to strongly depend on the details of the energy filtering process, a relationship which to date has not been satisfactorily explored. Here, we establish the conditions under which electron-hole separation in hot-carrier solar cells can occur reversibly, that is, at maximum energy conversion efficiency. We thus focus our analysis on the internal operation of the hot-carrier solar cell itself, and in this work do not consider the photon-mediated coupling to the sun. After deriving an expression for the voltage of a hot-carrier solar cell valid under conditions of both reversible and irreversible electrical operation, we identify separate contributions to the voltage from the thermoelectric effect and the photovoltaic effect. We find that, under specific conditions, the energy conversion efficiency of a hot-carrier solar cell can exceed the Carnot limit set by the intra-device temperature gradient alone, due to the additional contribution of the quasi-Fermi level splitting in the absorber. We also establish that the open-circuit voltage of a hot-carrier solar cell is not limited by the band gap of the absorber, due to the additional thermoelectric contribution to the voltage. Additionally, we find that a hot-carrier solar cell can be operated in reverse as a thermally driven solid-state light emitter. Our results help explore the fundamental limitations of hot-carrier solar cells, and provide a first step towards providing experimentalists with a guide to the optimal configuration of devices.

# 1. Introduction

In photovoltaic devices, photons with energy equal to or greater than the band gap of the absorber material are absorbed creating mobile electron-hole pairs, which are then separated to the negative and positive terminals of the device, respectively. In conventional, solid-state solar cells, carrier separation is achieved by differing carrier conductivities in the collectors adjacent to the absorber: n- and p- doped regions serve as electron and hole collectors, respectively, enabling electron (hole) transport to the negative (positive) terminal, while inhibiting hole (electron) transport to the negative (positive) terminal [1,2]. The open-circuit voltage of a single absorber solar cell is equal to the splitting of the quasi-Fermi levels (i.e. electrochemical potentials) [3] due to the non-equilibrium populations of electrons and holes in the absorber, and cannot exceed the absorber band gap [4] as population inversion and stimulated emission occur when the quasi-Fermi level splitting exceeds the band gap [5].

When a photon creates an electron-hole pair, any photon energy in excess of the band gap is divided between the electron and hole in a proportion that depends on the band structure. This can result in carriers with kinetic energy greatly in excess of the thermal energy, $E = 3/2\ kT$. In single absorber solar cells, electrons and holes lose this excess energy by inelastic carrier-phonon scattering before they are separated. This thermalization process represents a significant source of irreversible energy loss [6]. Hot-carrier solar cells are envisioned to extract energy from photogenerated electron-hole pairs before they cool down to the lattice temperature [7], thereby increasing the portion of the photon energy that can be extracted. Energy-filtering schemes have been proposed to selectively extract such high-energy carriers and utilize their energy [8,9].

Conceptually, substantial overlap exists between hot-carrier solar cells and nanoscale thermoelectric devices, which utilize temperature gradients to generate electric current: both aim to employ energy-filtering to extract energy from a temperature difference between charge-carrier populations in an optimal manner. In the field of thermoelectrics, it has proven very useful to analyze ideal energy filters based on quantum dots in order to understand thermodynamic limits. Specifically, quantum wells and quantum wires have been proposed as ideal thermoelectric devices [10,11], quantum dot heat engines have been analyzed under reversible operation [12] as well as in the maximum power regime [13-17], as have quantum well heat engines [18]. The open-circuit voltage of single quantum dots has also been studied experimentally [19,20] and quantum dot-like states in nanowires have been experimentally demonstrated to increase the thermoelectric power output [21].

Experimental work on hot-carrier solar cells to date has been divided into two generally isolated areas of investigation: first, an effort to slow the carrier cooling process by engineering absorber materials [22-27] and second, an effort to fabricate quantum wells and quantum dots and utilize them as energy selective filters [28-35]. Ultrafast hot-carrier charge separation has been investigated [36] and recently, some power generation in a hot-carrier solar cell based on an $Al_XGa_{1-X}As$ heterostructure was demonstrated [37].

Theoretical work on hot-carrier solar cells has so far been limited in scope. Specifically, in exploring the limits of the power conversion efficiency of hot-carrier solar cells to date [8,38-39], the existing literature simply assumes that hot-carrier extraction and energy conversion occurs reversibly at all device operating voltages. This assumption is problematic because reversible processes produce no power, and because we know from the literature on quantum heat engines [12-14] that reversibility is only achievable under very specific strong-coupling conditions [40].

In this work, we will more precisely consider the conditions under which electron-hole separation in hot-carrier solar cells can occur reversibly and we will quantify the increase in power production when hot-carrier solar cells are operated away from this point. In order to be able to analyze the internal, electrical operation of a hot-carrier solar cell, we explicitly exclude the thermodynamic coupling to the sun from the analysis in this paper. By analyzing the entropy generation during photogenerated electron-hole separation, we derive an expression for the voltage of a hot-carrier solar cell that is valid under conditions of both reversible and irreversible electrical operation, and show that it has separate contributions due to the thermoelectric effect and the photovoltaic effect. We establish that the maximum energy conversion efficiency (in the absence of power production) is not limited by the Carnot efficiency set by the temperature gradient within a hot-carrier solar cell, but can, under specific conditions, be higher due to the non-equilibrium provided by the quasi-Fermi level splitting in the absorber, which serves as a driving force in addition to the temperature gradient. Furthermore, we find that the open-circuit voltage of a hot-carrier solar cell is not limited by the band gap of the absorber, due to the thermoelectric contribution to the voltage. Additionally, we find that a hot-carrier solar cell can be operated in reverse as a thermally driven light emitter and we establish the conditions for its reversible operation.

The contribution of our work is to establish, for the first time, the operational point for reversible electron-hole separation in a hot-carrier solar cell, in analogy to the first works on quantum heat engines [12,13], thus establishing fundamental limits of performance. In addition, our results are also intended to serve as a guide to experimentalists who wish to design model systems that can test and explore these limits.

**2. Model system**

**2.1 Scope and overall approach**

A photovoltaic system can be considered to consist of three elements: (i) an illumination source (the sun), (ii) an absorber element, and (iii) electron and hole collectors. Two thermodynamic couplings connect these three elements: that between the sun and the absorber (mediated by photons), and that between the absorber and the collectors (mediated by charge-carriers and phonons).

The present study will focus on the electrical, charge-carrier mediated, thermodynamic coupling between the absorber and the collectors. We do not explicitly consider the details of the coupling between sun and absorber, because this requires a large number of material- and device-specific assumptions regarding the device geometry, the interaction between photons and carriers, the energy-dependent carrier relaxation rates (see also Sect. 7 below), and about phonon-mediated heat flow, which is also not explicitly included in the analysis. Instead, we will in the following describe the steady-state influence of the illumination source in terms of a resulting quasi-Fermi level splitting within the absorber, $\Delta\mu$ (a measure of the concentration of electron-hole pairs) and some absorber carrier temperature, $T_1$ (see also Fig. 1) bounded from above by the illuminating source temperature. This approach enables a transparent and general study of the conditions under which electron-hole separation in a hot-carrier solar cell can occur reversibly as function of the variable parameters $\Delta\mu$ and $T_1$, which describe the sun-absorber coupling, without clouding the analysis with specific assumptions about this coupling. Therefore, this work will contain no reference to the solar temperature. A limitation of this approach is that it doesn't allow us to calculate the power output under solar illumination, nor any solar power conversion efficiency, which both are descriptions of the complete three-element system. Rather, where we do compute efficiencies below, we will assume a monochromatic spectrum in order to enforce a strong coupling between the illumination source and absorber. The next step – beyond the scope of this work – will be an investigation of the maximum power limit of hot-carrier solar cells, a task that will require a full, self-consistent system analysis of the solar cell coupled to the sun, as we will discuss briefly in Section 7.

**2.2 Model details**

We consider a one-dimensional device, such as a heterostructure nanowire [20] containing two double-barrier heterostructures embedded within the intrinsic region of a p-i-n diode with a wide band gap (Fig. 1). The wide band gap of the p-type and n-type regions ensures that light absorption and electron-hole pair generation occurs only in the central absorber region (between the double-barriers), which has a narrow band gap $E_g$. On the left (right) of the absorber is a electron (hole) filter, that is, a resonant tunneling structure through which carriers are transmitted only at one specific resonant energy (any other resonances are assumed to be many thermal energies $kT$ away). The n-type and p-type collector regions for electrons and holes, respectively, are in Ohmic contact with an external circuit. The photo-generation of carriers in the absorber region results in a splitting of the electron and hole quasi-Fermi levels in the absorber by an amount $\Delta\mu$. Additionally, the device may have some voltage $V$ across it, that is, a difference $eV$ in the electrochemical potentials in the collector regions, where $e$ is the absolute value of the elementary charge.

The device is assumed to work in the following quasi-equilibrium regime. Arriving photons of energy $E_{ph} > E_g$ are absorbed and generate electron-hole pairs. We assume that electrons and holes retain a large fraction of the excess energy ($E_{ph}$ - $E_g$) amongst themselves by scattering with each other (electrons with electron, holes with holes, and electrons with holes) to establish a uniform carrier temperature $T_1$ that is higher (potentially much higher) than the lattice temperature. Essentially, this means that we assume that hot-carriers are extracted from the absorber on a time scale that is slow compared to that of inelastic carrier-carrier scattering, but fast compared to carrier-phonon scattering. Indeed, whereas relaxation and cooling times depend on many factors such as carrier concentration, material, confinement, and lattice temperature, photogenerated carriers thermalize amongst themselves typically in a few picoseconds whereas they cool to the lattice temperature in the time span hundreds of picoseconds at room temperature [41] and much slower at cryogenic temperatures. In comparison, the characteristic time for resonant tunneling through a double barrier structure (carrier extraction) is on the order of femtoseconds [42,43], and can be extended by tuning the size of the absorber region. Thus, there exists a time window for hot-carrier extraction that can be optimized by engineering the system parameters.

The resulting carrier temperature in the absorber (defined as $T_1$ in Fig. 1) can, in principle, be very high. For example, if a photon of energy 1 eV is absorbed in InAs ($E_g$ = 0.35 eV) and the excess energy ($E_{ph}$ - $E_g$) is divided equally between photogenerated electron and hole through electron-hole scattering and none of it is lost before extraction, each carrier will have a kinetic energy of 0.325 eV, corresponding to a temperature of 2514 K (as $E = 3/2\ kT$). In principle, even higher temperatures can be achieved with greater photon energies and smaller absorber bandgaps [44]. Experimentally, steady-state temperature differences between the lattice and photogenerated carrier distributions in excess of 100 K have been measured by continuous wave photoluminescence [26,27].

The essential thermodynamic parameters of the Figure 1 model system are defined in Figure 2. Once thermal equilibrium at $T_1$ is established within the carrier system in the absorber, electrons and holes therein can be characterized by their own quasi-Fermi levels $\mu_{n1}$ and $\mu_{p1}$ respectively [3,45], where we define $\Delta\mu = \mu_{n1} - \mu_{p1}$ (see Supplemental Information on details for how to calculate quasi-Fermi levels for a given bulk semiconductor and carrier concentration). The electron and hole energy filters are characterized by a single resonant energy level each, at energies $\varepsilon_n$ and $\varepsilon_p$ for electrons and holes, respectively, and we define $\Delta\varepsilon = \varepsilon_n - \varepsilon_p$. As electrons and holes are exchanged between the absorber and the collectors only at these energies, we consider the case of strong coupling between the absorber and the collectors, a thermodynamic condition in which heat flux is directly proportional to work-generating flux [40]. For the initial thermodynamic analysis, we assume infinitely sharp transmission resonances corresponding to relatively thick energy barriers [20]. For Landauer transport model calculations later in the paper, we will describe the transmission coefficient of the resonant energy levels using a Lorentzian, $\tau_n(E,V) = t_0/[1+\{(E-(\varepsilon_n-eV/2))/\delta\}^2]$, with an amplitude of $t_0 = 1$ and a full width at half maximum of $2\delta$. The two collector regions have equal temperature $T_2$ and are described by quasi-Fermi levels $\mu_{n2}$ and $\mu_{p2}$ respectively, and $\mu_{n2} - \mu_{p2} = eV$.

### 3. Reversible electron-hole separation in a hot-carrier solar cell

In the following section, we derive the conditions under which electron-hole separation in a hot-carrier solar cell as defined in Figures 1 and 2 can occur reversibly (without thermodynamic losses) by analyzing the generation of entropy during the extraction of an electron-hole pair from the absorber to the collectors, following an approach used to analyze quantum heat engines [12].

When an electron is extracted from the absorber through energy level $\varepsilon_n$, the change in entropy in the absorber is

$$\Delta S_{n1,ext} = \frac{-Q_{n1}}{T_1} = \frac{-(\varepsilon_n - \mu_{n1})}{T_1}. \quad (1)$$

The corresponding entropy change in the collector due to the injection of the electron is

$$\Delta S_{n2,inj} = \frac{Q_{n2}}{T_2} = \frac{\varepsilon_n - \mu_{n2}}{T_2}. \quad (2)$$

Similarly, when a hole is extracted from the absorber through energy level $\varepsilon_p$, the change in entropy in the absorber is

$$\Delta S_{p1,ext} = \frac{-Q_{p1}}{T_1} = \frac{-(\mu_{p1} - \varepsilon_p)}{T_1}, \quad (3)$$

and the corresponding entropy change in the collector due to the injection of the hole is

$$\Delta S_{p2,inj} = \frac{Q_{p2}}{T_2} = \frac{\mu_{p2} - \varepsilon_p}{T_2}. \quad (4)$$

The total change in system entropy due to the extraction and injection of the electron-hole pair is given by the sum of (1), (2), (3) and (4)

$$\Delta S = \Delta S_{n1,ext} + \Delta S_{n2,inj} + \Delta S_{p1,ext} + \Delta S_{p2,inj}$$
$$= \frac{\Delta \varepsilon - eV}{T_2} - \frac{\Delta \varepsilon - \Delta \mu}{T_1}. \quad (5)$$

Equation (5) is the first key result of this paper. It holds when the electron and hole energy filters are infinitely sharp $(\delta \to 0)$ and describes the overall increase in entropy upon separation of one electron-hole pair generated in the absorber. Importantly, the result shows that it is possible to define a working point where $\Delta S$ is zero, namely when

$$\frac{\Delta \varepsilon - \Delta \mu}{T_1} = \frac{\Delta \varepsilon - eV}{T_2}. \quad (6)$$

Under these conditions, an electron-hole pair can be extracted from the absorber and injected into the collectors of a hot-carrier solar cell as defined in Figures 1 and 2 reversibly, without thermodynamic losses. As this condition defines the case in which the photogenerated electron-hole separation process is reversible, it describes equally well the conditions under which the opposite process (the injection of an electron-hole pair into the absorber from the collectors) occurs reversibly.

This reversible condition described in equation (6) defines a state of energy-specific equilibrium that has been previously introduced in the context of quantum dot heat engines [12], reversible thermoelectric nanomaterials [46] and more generally, is obtained for strongly coupled thermodynamic systems [40]. To see this, consider that electrons in the absorber are in energy-specific equilibrium with the electrons in the electron collector when the arguments in the two Fermi-Dirac distributions are equal at the energy $\varepsilon_n$ where the two systems are in contact [12,46]

$$\frac{\varepsilon_n - \mu_{n1}}{T_1} = \frac{\varepsilon_n - \mu_{n2}}{T_2}, \quad (7)$$

a condition that can be obtained by setting the sum of (1) and (2) to zero. Similarly, the holes in the absorber are in energy-specific equilibrium with the hole collector when

$$\frac{\mu_{p1} - \varepsilon_p}{T_1} = \frac{\mu_{p2} - \varepsilon_p}{T_2}, \quad (8)$$

a condition that can be obtained by setting the sum of (3) and (4) to zero. At such points of energy-specific equilibrium as described by (7) and (8), there is no net charge current or heat current between the hot and cold carrier reservoirs (i.e. an open-circuit voltage condition is described). Summing (7) and (8), (6) is recovered.

Equation (6) can be graphically represented by a surface (Fig. 3) that shows the relationship between the model system thermodynamic parameters that results in zero entropy generation due to electron-hole separation. The zero entropy surface is divided into two sections by a black line that lies on the surface on the points at which $\Delta\mu = eV$. Along this line, $T_1 = T_2$. This line corresponds to the zero entropy operation point of a conventional (isothermal) single absorber solar cell: the open-circuit voltage point at which the quasi-Fermi level splitting at the contacts is equal to the quasi-Fermi level splitting in the absorber.

The regions on the zero entropy surface above and below the black line represent two different operational modes of the model system. Above the line, where $eV > \Delta\mu$ and $T_1 > T_2$, the model system operates as a hot-carrier solar cell: the absorption of light drives the system out of thermal and chemical equilibrium, creating temperature and quasi-Fermi level gradients that drive carriers into the collectors and through an external circuit. Below the line, where $eV < \Delta\mu$ and $T_1 < T_2$, the system is in the reverse mode of a hot-carrier solar cell, namely a thermally driven light emitter: an externally imposed thermal gradient between collectors and absorber drives carriers from the collectors into the absorber where they radiatively recombine, emitting light.

This is the second key result of this paper: a hot-carrier solar cell can be operated in reverse as a thermally driven light emitter and the point of zero entropy generation due to electron-hole separation of a hot-carrier solar cell also describes the zero entropy operation point of a thermally driven light emitter.

The parameters used in Fig. 3, namely $T_1$, $T_2$, $\Delta\mu$, and $\Delta\varepsilon$, are, in principle, experimentally controllable. Specifically, $\Delta\varepsilon$ is controllable by band engineering of the experimental device through choice of materials and layer thicknesses. $\Delta\mu$ is controlled by the photon flux of the illumination source and the recombination rates within the absorber. $T_1$ is dependent upon the average illuminating photon energy, the absorber band gap and the carrier-phonon interaction rates within the absorber. $T_2$ is the steady-state temperature of the crystal lattice in the collectors, and depends on the device temperature and the phonon-mediated heat flow from the absorber. As described in Section 2.1, we treat $\Delta\mu$, $T_1$ and $T_2$ as tunable variables so that we can understand their role for the thermodynamics of a hot-carrier solar cell, even though in practice they will be interdependent, and only a fraction of the parameter space in Fig. 3 will be experimentally accessible for a given device and illumination source.

**4. Voltage of a hot-carrier solar cell**

Equation (5) can be rearranged to describe the voltage of the model hot-carrier solar cell:

$$eV = \Delta\varepsilon\left(1 - \frac{T_2}{T_1}\right) + \Delta\mu\frac{T_2}{T_1} - T_2\Delta S. \quad (10)$$

Equation (10) is at its core an expression familiar from the hot-carrier solar cell literature (see for example Eq. 3 in [38]), with the very important addition that it includes an additional, electrical entropic loss term. In other words, Eq. (10) describes the reduction of the voltage due to entropy generation – a requirement for power production and irreversible operation – that, to the best of our knowledge, has not been previously accounted for in the hot-carrier solar cell literature.

By utilizing an expression for the amount of heat which is extracted from the absorber when an electron-hole pair is extracted

$$Q_1 = Q_{n1} + Q_{p1} = (\varepsilon_n - \mu_{n1}) + (\mu_{p1} - \varepsilon_p) = \Delta\varepsilon - \Delta\mu, \quad (11)$$

(10) can be rearranged into a more elucidating form, which highlights the way in which the two non-equilibriums present in a hot-carrier solar cell (i.e. the temperature difference $T_1 - T_2$ and the quasi-Fermi level splitting $\Delta\mu$) contribute to an electron-hole pair's ability to do work:

$$eV = Q_1\eta_{Carnot} + \Delta\mu - T_2\Delta S, \quad (12)$$

where $\eta_{Carnot} = (1 - T_2/T_1)$ is the intra-device Carnot efficiency. The three terms in (12) offer an intuitive picture: the first term describes an ideal heat engine operating at Carnot efficiency as described in [12-14] and constitutes the thermoelectric contribution to the voltage (i.e. it describes a conversion of heat into voltage). The second term describes the photovoltaic contribution to the voltage, namely the work attainable due to the free energy of the electron-hole pair: the difference in the quasi-Fermi levels of the electrons and the holes in the absorber. Finally, the third term quantifies the reduction in energy conversion performance when the engine is operated away from the zero entropy condition.

It is illustrative to consider two limiting cases of (12). First, we consider the case without a temperature gradient ($T_1 = T_2$ and thus $\eta_{\text{Carnot}} = 0$), that is, that of a conventional solar cell. Equation (12) then reduces to an expression similar to the textbook description of the voltage provided by a conventional, single absorber solar cell [3]

$$eV = \Delta\mu - T\Delta S. \qquad (13)$$

The form presented here varies from that in [3] through the introduction of the entropic loss term, which quantifies the consequence of operating a solar cell away from the zero entropy condition – a requirement to produce power. One way of looking at this is to realize that the use of spatially varying quasi-Fermi levels (e.g. a case in which the quasi-Fermi level splitting at the contacts is less than that in the absorber, $eV < \Delta\mu$) is a requirement for driving current when temperature is uniform [1,2]. As carriers move from the absorber to the collectors and contacts driven by the quasi-Fermi level gradient, they increase in entropy, which reduces their ability to do work. In this way, the introduction of the entropic loss term thermodynamically quantifies the difference between the open-circuit voltage (i.e. $\Delta\mu$) and the voltage at other points on the current-voltage curve such as the voltage at maximum power, and provides a thermodynamic explanation for why the fill factor of solar cells must be less than one.

As a second limit, we consider Eq. (12) for the case $\Delta\mu = 0$, that is, that of an electron-hole heat engine as described in [14]:

$$eV = \Delta\varepsilon\left(1 - \frac{T_2}{T_1}\right) - T_2\Delta S \qquad (14)$$

(14) appears as Eq. 7 in [14] as a result describing the case when Carnot efficiency is achieved: $\Delta S$ is zero and current through the system is zero, making the expression a description the open-circuit voltage of the system. In the form presented here, (15) describes a more general case applicable to all operating points of an ideal electron-hole heat engine. The addition of the entropic loss term indicates that at all operating points on the device current-voltage curve (excepting the open-circuit voltage point when $\delta \to 0$) some amount of entropy is generated by charge-carrier movement through the system, which decreases the amount of work which the moving charge-carriers can do on a load.

## 5. Energy conversion efficiency

To calculate the maximum energy conversion efficiency of our model system, we consider the case where a device described by a given $T_1$ and $\Delta \mu$, is illuminated by a monochromatic source with photon energy $E_{ph}$, and is operated under conditions where $\Delta S = 0$, that is, somewhere along the zero entropy surface illustrated in Figure 3. When $\Delta S = 0$, the device is operating at its maximum open-circuit voltage at which point there is zero power production, equivalent to a quantum-dot heat engine at Carnot efficiency [12-14]. Dividing the output open-circuit voltage by the input photon energy gives the efficiency at which the electromagnetic energy of a photon is converted into the electrochemical potential energy of an electron-hole pair at this point. This yields the maximum energy conversion efficiency

$$\eta_{HCSC} = \frac{eV_{OC}}{E_{ph}} = \frac{(Q_1 \eta_{Carnot} + \Delta \mu)}{E_{ph}}. \quad (15)$$

Equation (15) is meaningful only for $E_{ph} \geq \Delta \varepsilon$ to ensure that the energy available for extraction from the electron-hole system doesn't exceed the energy available from the photon.

Equation (15) highlights the third key result of this paper: under specific conditions, the energy conversion efficiency of an ideal hot-carrier solar cell can exceed the Carnot limit set by the intra-device temperature gradient due to the additional contribution of the quasi-Fermi level splitting in the absorber. This is illustrated in Figure 4. Additionally, while $\Delta \mu$ is bounded by the band gap of the absorber [4,5],

$Q_1\eta_{Carnot}$, is not, enabling the open-circuit voltage of a hot-carrier solar cell to exceed the band gap of the absorber through the conversion of carrier heat into voltage in the manner of a thermoelectric device.

The fact that a hot-carrier solar cell energy conversion efficiency, $\eta_{HCSC}$, as discussed here can exceed the intra-device Carnot efficiency, $\eta_{Carnot}$, does not violate any laws of thermodynamics. As stated previously, it has been assumed that the temperature of the illumination source is greater than that of the absorber. Thus, there is a larger Carnot efficiency, which characterizes the complete, three element, dual-coupled system. This complete-system Carnot efficiency (involving the temperature of the sun) cannot be exceeded by $\eta_{HCSC}$.

As shown in Figure 4a, for a set value of $T_2$, and when $\Delta\varepsilon = E_{ph}$ and $\Delta\mu > 0$, the $\eta_{HCSC}$ exceeds $\eta_{Carnot}$ for all values $T_1 \geq T_2$ by an amount that decreases as the $T_2/T_1$ ratio and the $\Delta\mu/E_{ph}$ ratio decrease (blue and red curves). When $\Delta\varepsilon < E_{ph}$ and $\Delta\mu > 0$, there is a range of $T_1$ in which $\eta_{HCSC}$ exceeds $\eta_{Carnot}$ and a range in which it does not (green curve in Figure 4a).

Another way to visualize the relationship between $\eta_{HCSC}$ and $\eta_{Carnot}$ is to choose a set $T_2/T_1$ ratio and to plot the efficiency of the device according to (15) as a function of the $\Delta\varepsilon/E_{ph}$ ratio for different values of $\Delta\mu$. This is done in Figure 4b for $T_2 = 300$ K and $T_1 = 600$ K and illustrates three messages. Firstly, for a given $\Delta\mu > 0$ value and $T_2/T_1$, there is a $\Delta\varepsilon/E_{ph}$ ratio range in which the Carnot limit is exceeded and a range in which it is not. This case is illustrated in the red curve in Figure 4b for the case of $\Delta\mu = 0.5$ eV. Secondly, Figure 4b shows that $\eta_{Carnot}$ can be achieved in a specific case when $\Delta\mu = 0$: when the energy available for input into the absorber and extraction from the absorber are exactly equal, $\Delta\varepsilon = E_{ph}$ (green curve). Finally, Figure 4b shows that the Carnot efficiency cannot be achieved when $\Delta\mu < 0$ (blue curve) (see Supplemental Information for details on this condition).

When the temperature is uniform, the energy conversion efficiency of a conventional solar cell is recovered, which (obviously) exceeds the intra-device Carnot limit set by the temperature difference between the absorber and the collectors (i.e. $\eta_{\text{Carnot}} = 0$, since $T_1 = T_2$) is for all positive values of $\Delta\mu$. This efficiency depends solely upon the $\Delta\mu/E_{\text{ph}}$ ratio and is plotted in Figure 4c for the case of $\Delta\mu = 0.5$ eV. The case of such a conventional, single absorber solar cell and its thermodynamic coupling to an illumination source has been considered in [47] using a master equation formulation for driven open systems and (15) is identical to Eq. 9 in [47] for $T_1 = T_2 = T$ and when non-radiative recombination is absent.

## 6. Landauer model and power conversion efficiency

Equation (6) shows that there is one value of $eV$ (i.e. one operating voltage) at which point $\Delta S$ will be zero for a device at $T_2$ with a set $\Delta\varepsilon$ (defined by the device design) and that is further characterized by some $\Delta\mu$ and $T_1$ (a function of the illumination spectrum and intensity). This prediction can be verified and visualized by placing the given parameters into a Landauer transport model (see Supplemental Information for details) and calculating the current-voltage curve and the entropy generation at each point on this curve. Specifically, Figures 5a and 5c illustrate that the zero entropy point corresponds to the open-circuit voltage of the device and that the open-circuit voltage occurs at the voltages prescribed by (6) when $\delta$ is very narrow. Figures 5a and 5c also illustrate that the short-circuit current increases with increasing $\Delta\mu$.

The Landauer transport model allows us to explore the effect of non-ideal energy filters with a finite width $\delta$ (a prerequisite for achieving power production). In such cases, a hot-carrier solar cell power conversion efficiency expression can be defined by dividing the generated electrical power by the rate at which monochromatic photons must be supplied in order to sustain the current being extracted from the absorber (see Supplemental Information Eq. S17). Figure 5b and Figure 5d illustrate that increasing $\delta$ results in decreased maximum power conversion efficiency and open-circuit voltage, but increased maximum power.

## 7. Conclusion

In conclusion, we have identified the conditions under which electron-hole separation in hot-carrier solar cells can occur reversibly and we have derived an expression for the voltage provided by a hot-carrier solar cell that is valid under conditions of both reversible and irreversible electrical operation. Importantly, in full analogy to quantum heat engines, reversible electrical operation is possible only at one specific operation point for a given device (i.e. a set $\Delta\varepsilon$) operating at a given temperature, $T_2$, and characterized by a certain degree of non-equilibrium (i.e. set values of $\Delta\mu$ and $T_1$). At this point, under particular conditions, the energy conversion efficiency of a hot-carrier solar cell can exceed the Carnot limit set by the intra-device temperature gradient due to the quasi-Fermi level splitting in the absorber, which serves as a non-equilibrium driving force in addition to the temperature difference. Additionally, we find that the open-circuit voltage of a hot-carrier solar cell is not limited by the band gap of the absorber due to the thermoelectric contribution to the voltage.

It is worth noting that, if operated at the energy-specific equilibrium, open-circuit voltage point (6), where entropy generation due to electron-hole separation is zero, the lattice of a hot-carrier solar cell will reach a high steady-state temperature, because photogenerated carriers will lose some of their excess energy to phonons before recombining, and no charge-carriers or electronic heat will be extracted from the absorber to the collectors. The steady-state operation in this regime thus involves phonon-mediated heat flow (with corresponding phononic entropy production) from absorber to collectors that is not included in our analysis (see Section 2.1).

We also found that the model system and thermodynamics that describe a hot carrier solar cell and its point of zero entropy generation due to electron-hole separation also describe a thermally driven light emitter and its zero entropy operation point. Such a thermally driven light emitter would extract heat from the lattice and dump it into a photon bath acting as a luminescent cooler, which, in principle, may be used to produce light from heat, or to use photons as a way of extracting excess heat, for example from hot spots in integrated circuits. The detailed operation of such a device, including its limiting efficiency, will be a topic for future investigation.

Our work provides a number of useful "take-aways" for experimentalists. Relevant to the field of photovoltaics, we defined a model device structure and we described the thermodynamics for how to obtain an open-circuit voltage that is not limited by the band gap of the absorbing material, a fundamental limitation in conventional, uniform temperature solar cells [4,5]. In the field of thermoelectrics, we showed that, by illuminating a device from a source with a temperature in excess of the hot temperature within the device, the Carnot limit set by the intra-device temperature gradient can be exceeded due to the additional non-equilibrium driving force of the split quasi-Fermi levels. Finally, we showed how to obtain LED-like light emission from a semiconductor device without any electrical biasing: by employing a temperature gradient between device regions to drive radiatively recombining thermocurrents. To our knowledge, such a combined treatment of solar cells, thermoelectrics and solid-state light emitters has in this form not been undertaken before.

While in this work, $T_1$ and $\Delta\mu$ have been taken as variable parameters in order to focus on the thermodynamics of photogenerated electron-hole separation, in future work, they will need to be calculated self-consistently based upon energy balance and particle balance models taking into account the complete three element, dual-coupled system, including illumination spectrum and intensity, absorber band gap, relaxation times and the rates of electron-hole pair recombination and extraction. Such a self-consistent treatment will allow the calculation of the complete system energy and power conversion efficiencies. Furthermore, for nanoscale realizations of hot-carrier solar cells, it may be of interest to explore the role of fluctuations, for example that of temperature fluctuations in the absorber region due to stochastic carrier extraction and relaxation events [48].

It is also interesting to note that hot-carrier solar cells exhibit a number of characteristics that make them an ideal thermoelectric system for experimental investigation. Firstly, very high temperature gradients are possible: steady-state temperature differences between the lattice and photogenerated carrier distributions in excess of 100 K have been measured by continuous wave photoluminescence in recent studies [26,27] and much higher temperatures are theoretically achievable [44]. Secondly, heat losses due lattice thermal conduction – a major challenge of traditional thermoelectrics – is avoided by the separation of photogenerated carrier prior to their thermalization with the lattice, and in the ideal case, the lattice temperature remains uniform throughout the device. This may provide a novel path to achieving high-efficiency thermoelectric devices.

**Acknowledgements**

Financially supported by a Solander Fellowship, the Swedish Energy Agency (grant no. 38331-1), the Swedish Research Council, by NanoLund, the Center for Nanoscience at Lund University, and the Australian Government through the Australian Renewable Energy Agency (ARENA).

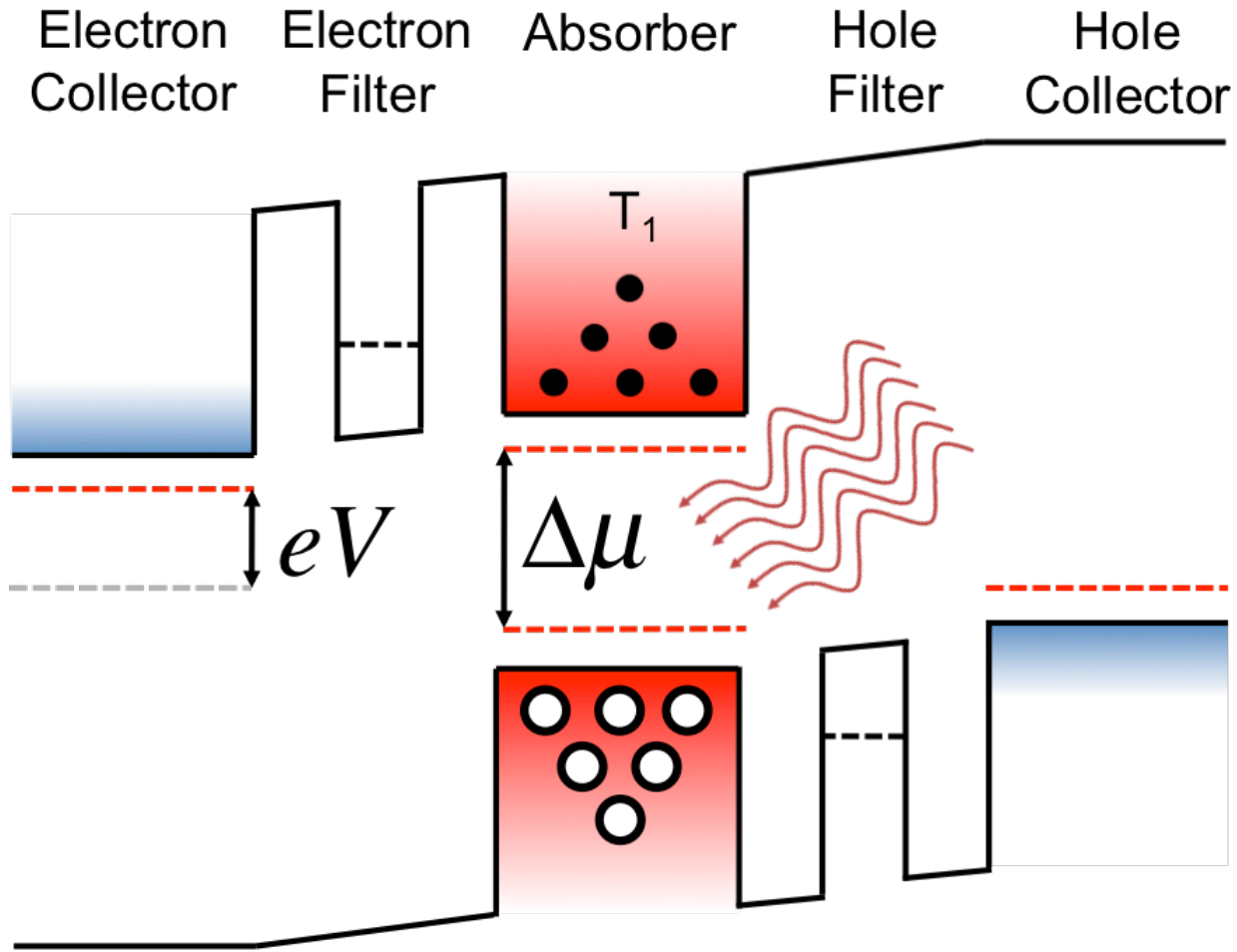

Figure 1: Model system for a hot-carrier solar cell defining five device regions including absorber and filters embedded in the intrinsic region of a wide bandgap p-i-n diode. Indicated on the figure are quasi-Fermi levels (dashed red lines) and energy levels of carrier exchange (dashed black lines).

Figure 2: Hot-carrier solar cell diagram defining thermodynamic parameters (see main text for a description).

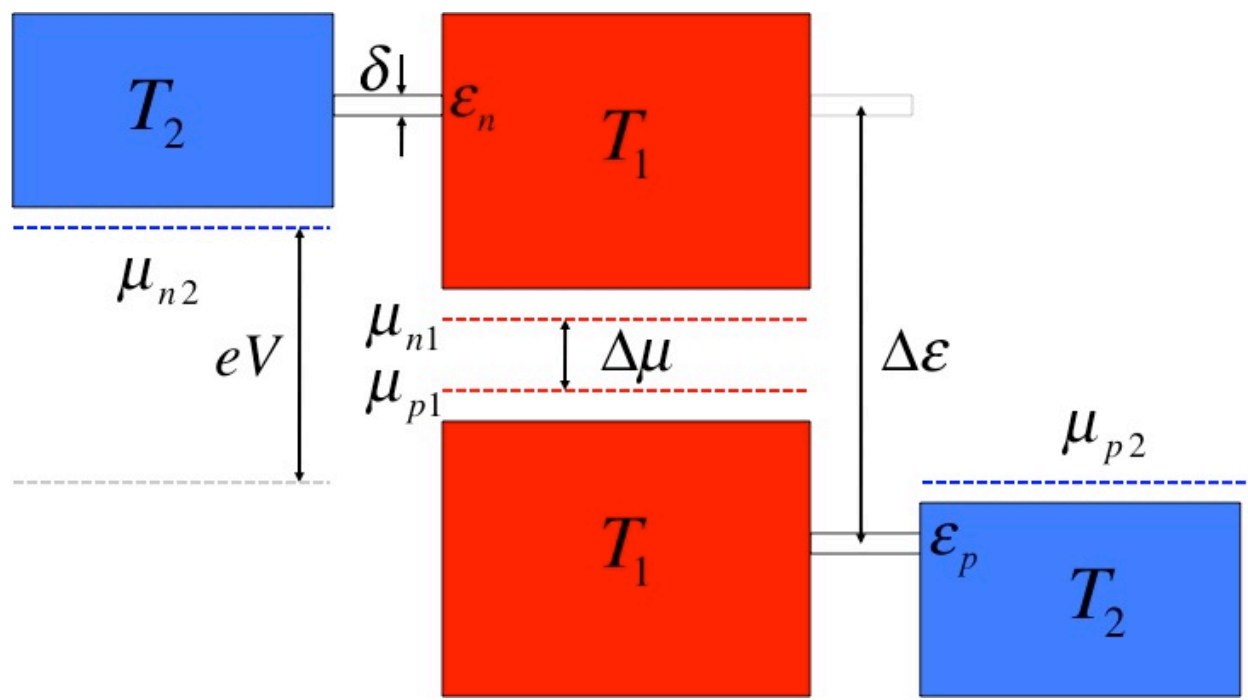

Figure 3: Surface along which $\Delta S = 0$ computed from equation (6) showing the relationship between thermodynamic parameters that results in zero entropy generation due to electron-hole separation. $T_2$ is taken to be 300 K and $T_1$ is calculated for $(\Delta\varepsilon - \Delta\mu)$ and $(\Delta\varepsilon - eV)$ from 0 to 1 eV using (6). The black line indicates the condition $\Delta\mu = eV$. Located above this line are the operation points of a hot-carrier solar cell. Located below the line are the operation points of a thermally driven light emitter.

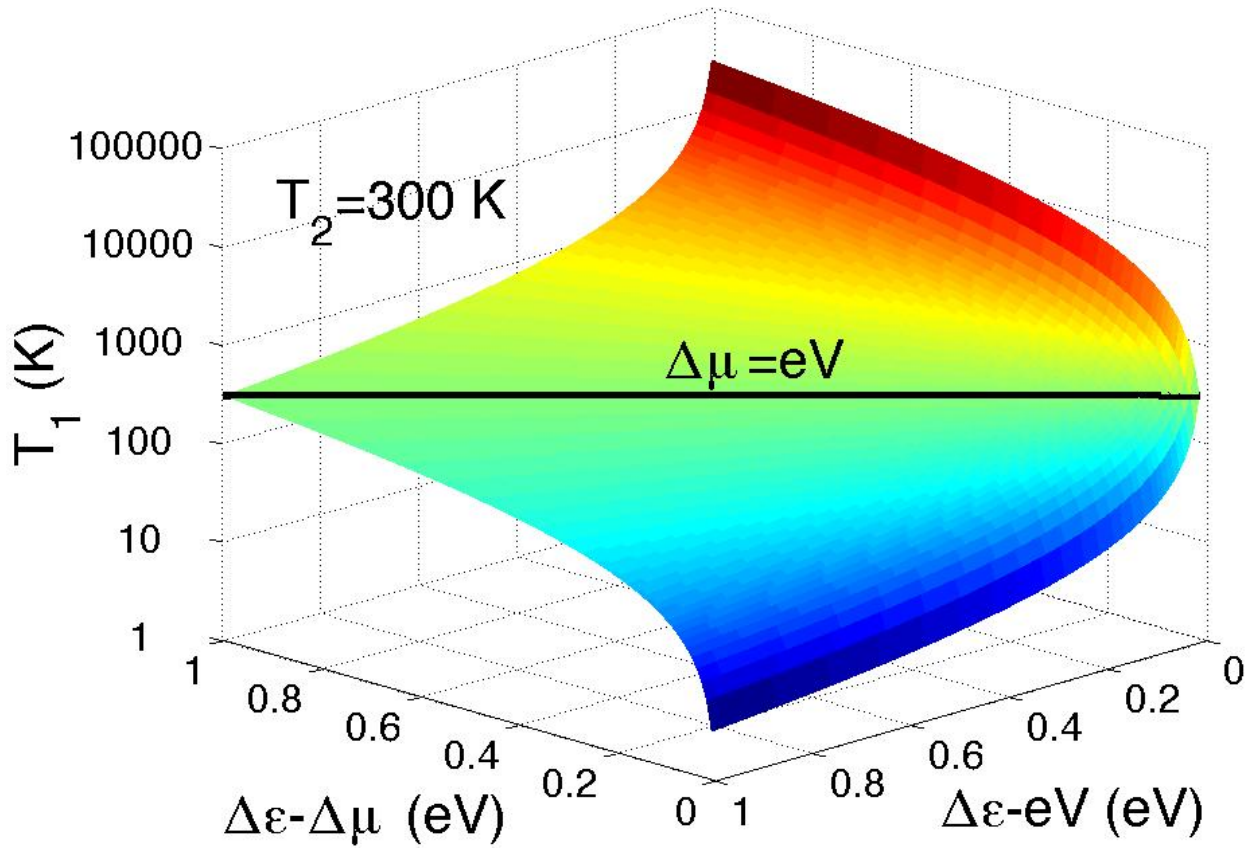

Figure 4: Maximum energy conversion efficiency of a hot-carrier solar cell (equation 15).

a) $\eta_{HCSC}$ as a function of hot-carrier temperature, $T_1$, for $T_2 = 300$ K and $\Delta\mu = 0.5$ eV for three different energy extraction and input scenarios: $\Delta\varepsilon = E_{ph} = 1$ eV (blue), $\Delta\varepsilon = E_{ph} = 3$ eV (red) and $\Delta\varepsilon = 2$ eV, $E_{ph} = 3$ eV (green).

b) $\eta_{HCSC}$ for $T_2 = 300$ K and $T_1 = 600$ K as a function of $\Delta\varepsilon/E_{ph}$ for three different values of $\Delta\mu$: $\Delta\mu = -0.5$ eV (cyan), $\Delta\mu = 0$ eV (orange), and $\Delta\mu = 0.5$ eV (magenta).

c) Energy conversion efficiency of a conventional single absorber solar cell ($T_1 = T_2$ in equation 15) with an open-circuit voltage of 0.5 V as a function of monochromatic illumination source photon energy, $E_{ph}$ (see main text for discussion).

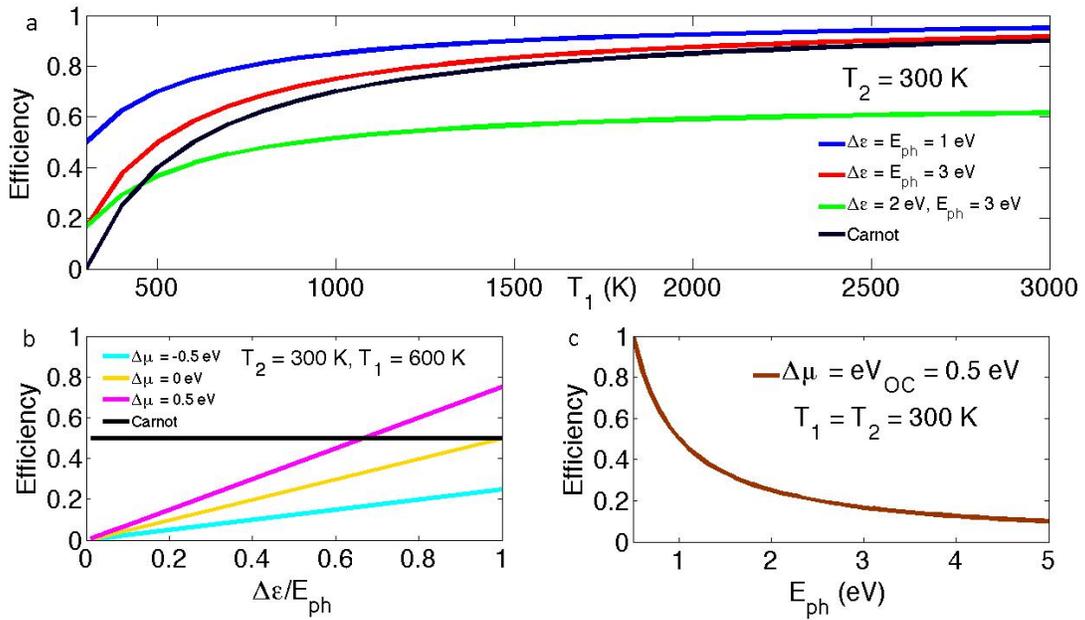

Figure 5: Operation of a hot-carrier solar cell with finite $\delta$.

a) Current-voltage curves computed from a Landauer model hot-carrier solar cell with $T_2=300$ K, $T_1 = 600$ K, $\Delta\varepsilon = 0.75$ eV and $\delta = 10^{-4}$ meV for three different values of $\Delta\mu$: $\Delta\mu = 0.1$ eV (green), $\Delta\mu = 0$ eV (red), and $\Delta\mu = -0.1$ eV (blue). The expected zero entropy operating points from (6) are $eV = 0.425$ eV, $eV = 0.375$ eV, and $eV = 0.325$ eV, respectively.

b) Power conversion efficiency curves (see Supplemental Information Eq. S17) computed from a Landauer model with $T_2 = 300$ K, $T_1 = 600$ K, $\Delta\varepsilon = 0.75$ eV and $\Delta\mu = 0.1$ eV for four values of $\delta$: $\delta = 10^{-4}$ meV (black), $\delta = 10^{-1}$ meV (magenta), $\delta = 1$ meV (orange), $\delta = 10$ meV (cyan).

c) Entropy-voltage curves corresponding to the current-voltage curves plotted in Figure 5a. The expected zero entropy operating points from (6) for the cases computed are $eV = 0.425$ eV (green), $eV = 0.375$ eV (red), and $eV = 0.325$ eV (blue), respectively.

d) Power-voltage curves corresponding to the power conversion efficiency curves plotted in Figure 5b.

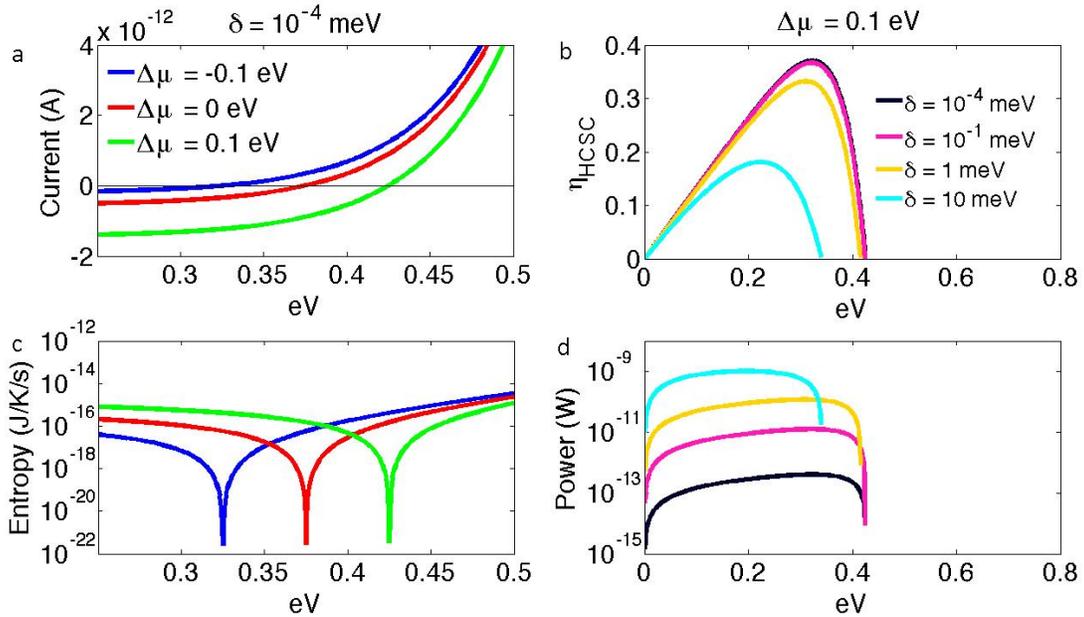

Supplemental Information

## 1. Calculation of electron and hole quasi-Fermi levels

In the following, we introduce the notion of quasi-Fermi levels following references [3,45]. The concentration of electrons in the conduction band of a bulk semiconductor is given by

$$n = \int_{E_c}^{\infty} D_c(E) f_n(E) dE$$
$$= \int_{E_c}^{\infty} \left( 4\pi \left( \frac{2m_n^*}{h^2} \right)^{3/2} \sqrt{E - E_c} \right) \left( \exp\left( \frac{E - \mu_n}{kT} \right) + 1 \right)^{-1} dE, \quad \text{(S1)}$$

where $D_C(E)$ is the density of states in the conduction band and $f_n(E)$ is the distribution function of the electrons in the conduction band, in this case, a Fermi-Dirac distribution. To determine $\mu_n$, this equation must be solved numerically unless $\mu_n < E_c - 3kT$ in which case the "+1" in the denominator may be ignored and the concentration of electrons can be expressed as

$$n = 2 \left( \frac{2\pi m_n^* kT}{h^2} \right)^{3/2} \exp\left( -\frac{E_c - \mu_n}{kT} \right) = N_C \exp\left( -\frac{E_c - \mu_n}{kT} \right), \quad \text{(S2)}$$

where $N_C$ is the effective density of states in the conduction band. Then, the electron quasi-Fermi level can be solved for as

$$\mu_n = E_c - kT \ln\left( \frac{N_C}{n} \right). \quad \text{(S3)}$$

Similarly, for holes

$$p = \int_{-\infty}^{E_v} D_v(E) f_p(E) dE$$
$$= \int_{-\infty}^{E_v} \left( 4\pi \left( \frac{2m_p^*}{h^2} \right)^{3/2} \sqrt{E_v - E} \right) \left( \exp\left( \frac{\mu_p - E}{kT} \right) + 1 \right)^{-1} dE, \quad \text{(S4)}$$

where $D_v(E)$ is the density of states in the valence band and $f_p(E)$ is the distribution function of the holes in the valence band, in this case, a Fermi-Dirac distribution. To determine $\mu_p$, this equation must be solved numerically unless $\mu_p > E_v + 3kT$ in which case the "+1" in the denominator may be ignored and the concentration of electrons can be expressed as

$$p = 2\left(\frac{2\pi m_p^* kT}{h^2}\right)^{3/2} \exp\left(-\frac{\mu_p - E_v}{kT}\right) = N_V \exp\left(-\frac{\mu_p - E_v}{kT}\right), \quad (S5)$$

where $N_V$ is the effective density of states in the valence band. Then, the hole quasi-Fermi level can be solved for as

$$\mu_p = E_v + kT \ln\left(\frac{N_V}{p}\right). \quad (S6)$$

## 2. Carrier concentration, temperature and negative quasi-Fermi level splitting

Using the expressions given above, we can explore the relationship between carrier concentration, temperature and quasi-Fermi levels and show that negative quasi-Fermi level splitting can occur at elevated carrier temperatures [38]. Let us take the material to be GaAs in order to specify the effective masses of the electrons and holes and the band edges. We can then solve for the quasi-Fermi level of the electrons and holes across a range of temperatures for selected concentrations. The results of these calculations are plotted in Figure S1 below for concentrations of 1e16 cm$^{-3}$ and 1e17 cm$^{-3}$ in the temperature range from 300 K to 3000 K.

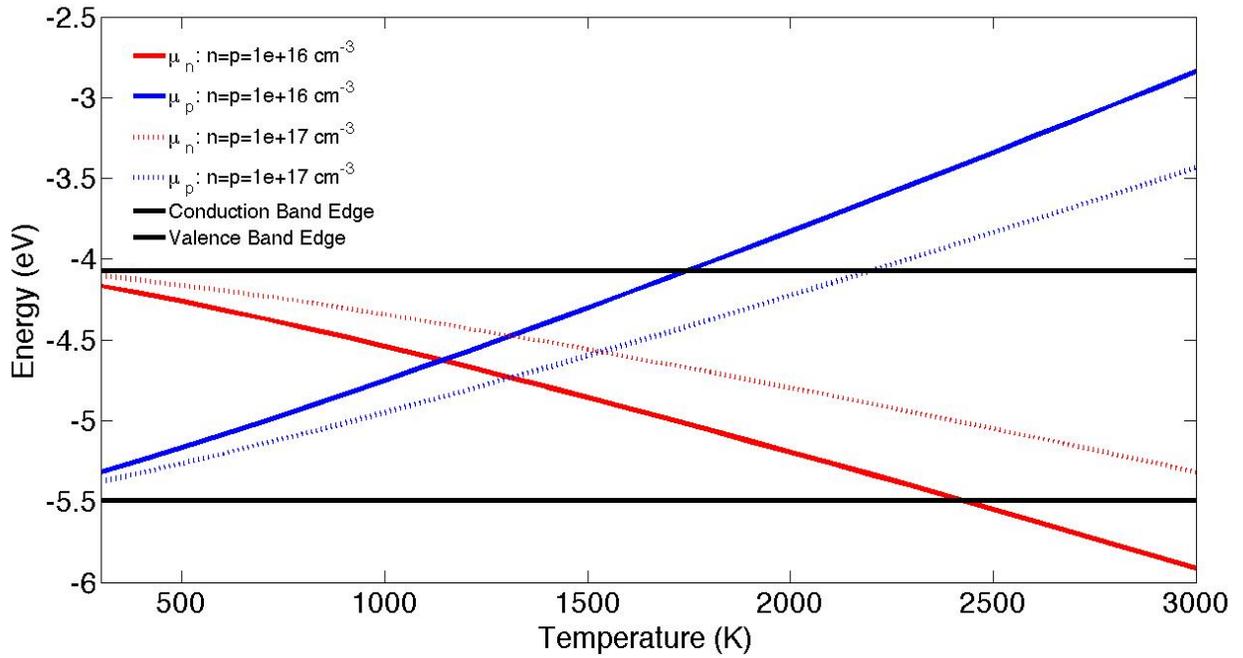

Figure S1: Electron and hole quasi-Fermi levels in GaAs as a function of temperature for two different non-equilibrium carrier concentrations.

The results show that with constant carrier concentration, as the temperature increases, the electron quasi-Fermi level decreases and the hole quasi-Fermi level increases. There is a concentration dependent crossing point at which the electron and hole quasi-Fermi levels are equal and beyond which $\Delta\mu = \mu_n - \mu_p < 0$. As the electrons and holes in hot-carrier solar cells may be at these extremely high temperatures, the consideration of the $\Delta\mu < 0$ case is relevant when analyzing such devices.

**3. Landauer transport model of a hot-carrier solar cell**

Data in Figure 5a were calculated using the following Landauer transport model

$$I_1(V) = -2\frac{e}{h}\int_{E_c}^{\infty}\tau_n(E,V)f_{n1}(E,V)dE \quad (S7)$$

$$I_2(V) = 2\frac{e}{h}\int_{E_c}^{\infty} \tau_n(E,V) f_{n2}(E) dE \qquad (S8)$$

$$I(V) = I_2(V) + I_1(V) \qquad (S9)$$

$$\tau_n(E,V) = 1/\left(1 + \left(E - \left(\varepsilon_n - eV/2\right)\right)/\delta\right)^2\right) \qquad (S10)$$

$$f_{n2}(E) = \left[\exp\left(\frac{E - \mu_{n2}}{kT_2}\right) + 1\right]^{-1} \qquad (S11)$$

$$f_{n1}(E,V) = \left[\exp\left(\frac{E - (\mu_{n1} - eV/2)}{kT_1}\right) + 1\right]^{-1} \qquad (S12)$$

Data in Figure 5c were calculated using the following extensions to the Landauer transport model

$$\dot{Q}_1(V) = 2\left[\frac{2}{h}\int_{E_c}^{\infty} \tau_n(E,V)\left(E - (\mu_{n1} - eV/2)\right)\left(f_{n2}(E) - f_{n1}(E,V)\right) dE\right] \qquad (S13)$$

$$\dot{Q}_2(V) = 2\left[-\frac{2}{h}\int_{E_c}^{\infty} \tau(E,V)(E - \mu_{n2})\left(f_{n2}(E) - f_{n1}(E,V)\right) dE\right] \qquad (S14)$$

$$\dot{S}(V) = \frac{\dot{Q}_1(V)}{T_1} + \frac{\dot{Q}_2(V)}{T_2} \qquad (S15)$$

Data in Figure 5d and Figure 5b were calculated using the following power and power conversion efficiency expressions:

$$P(V) = I(V)V \qquad (S16)$$

$$\eta_{HCSC}(V) = \frac{P(V)}{I_1(V)E_{ph}} \qquad (S17)$$

The power conversion efficiency expression states that the power input into the hot-carrier solar cell is the minimum monochromatic photon flux necessary to sustain the current extracted from the absorber at each voltage point on the current-voltage curve. This expression assumes that each absorbed photon generates one electron-hole pair that contributes to this extracted current, $I_1$. Additionally, in order to ensure that the energy available for extraction from the absorber electron-hole system doesn't exceed the energy available to be supplied by a photon, $E_{ph}$ is taken to be equal to $\Delta\varepsilon$.